\documentclass[%
aip,
amsmath,amssymb,
reprint,%
]{revtex4-1}

\usepackage{amssymb,amsmath,graphicx,amsthm,wrapfig,epstopdf,esint,afterpage,marginnote,afterpage,lipsum,color,tabularx}
\usepackage{graphics,graphicx}
\usepackage{dcolumn,bm}
\usepackage[dvipsnames]{xcolor}
\usepackage{natbib}
\usepackage[version=4]{mhchem}
\usepackage{color,sidecap,fancyhdr}
\usepackage{hyperref}
\usepackage{cleveref}

\newcommand{\eps}{\epsilon}
\newcommand{\braa}[1]{\left(#1\right)}
\newcommand{\Bra}[1]{\left[#1\right]}

\usepackage{mathtools, etoolbox}

\DeclarePairedDelimiterX\braket[2]{\langle}{\rangle}{#1 \delimsize\vert #2}

\newcommand{\colvec}[2]{
\begin{bmatrix}
#1\\
#2
\end{bmatrix}
}

\newcommand{\partiald}[2]{
\frac{\partial #1}{\partial #2}
}

\newcommand{\doublepartiald}[2]{
\frac{\partial^2 #1}{\partial #2^2}
}

\begin{document}

\title{Comb-like Turing patterns embedded in Hopf oscillations: Spatially localized states outside the 2:1 frequency locked region}

\author{Paulino Monroy Castillero}
\email{espaulino@gmail.com}
 \affiliation{Instituto de Ciencias F\'isicas, Universidad Nacional Aut\'onoma de M\'exico, Cuernavaca 62210, Mexico}
\author{Arik Yochelis}%
 \email{yochelis@bgu.ac.il}
\affiliation{%
 Department of Solar Energy and Environmental Physics, Swiss Institute for Dryland Environmental and Energy Research, Blaustein Institutes for Desert Research (BIDR), Ben-Gurion University of the Negev, Sede Boqer Campus, 8499000 Midreshet Ben-Gurion, Israel
}%

\date{\today}

\begin{abstract}
	A generic distinct mechanism for the emergence of spatially localized states embedded in an oscillatory background is demonstrated by using 2:1 frequency locking oscillatory system. The localization is of Turing type and appears in two space dimensions as a comb-like state in either $\pi$ phase shifted Hopf oscillations or inside a spiral core. Specifically, the localized states appear in absence of the well known flip-flop dynamics (associated with collapsed homoclinic snaking) that is known to arise in the vicinity of Hopf-Turing bifurcation in one space dimension. Derivation and analysis of three Hopf-Turing amplitude equations in two space dimensions reveals a local dynamics pinning mechanism for Hopf fronts, which in turn allows the emergence of perpendicular (to the Hopf front) Turing states. The results are shown to agree well with the comb-like core size that forms inside spiral waves. In the context of 2:1 resonance, these localized states form outside the 2:1 resonance region and thus extend the frequency locking domain for spatially extended media, such as periodically driven Belousov-Zhabotinsky chemical reactions. Implications to chlorite-iodide-malonic-acid and shaken granular media are also addressed.
\end{abstract}

\maketitle

\textbf{Experiments with the oscillatory chlorite-iodide-malonic-acid (CIMA) chemical reaction have demonstrated that spiral waves can exhibit a finite size stationary core, a.k.a. dual-mode spiral waves. The behavior had been attributed to a competition between the coexisting oscillatory (Hopf) and stationary periodic (Turing) instabilities through analysis in one-space dimension (1D). Specifically, localized stationary solutions have shown to emerge in between $\pi-$shifted oscillations and thus, assumed to explain the spiral core where the amplitude of oscillations vanishes. Yet, numerical simulations indicate that spatially localized comb-like states in 2D form outside the coexistence region that is obtained in 1D. Consequently, a distinct mechanism is derived via a weakly nonlinear analysis near the Hopf-Truing bifurcation in 2D and shown to well agree with numerical simulations. Moreover, the results are discussed in the context of 2:1 frequency locking and show that resonant localized patterns extend the standard frequency locking region. Consequently, the study suggests distinct control and design features to spatially extended oscillatory systems.}

\section{Introduction}\label{intro}

Chemical reactions are frequently being used as case models to elucidate generic and rich mechanisms of spatiotemporal dynamics, such as the Turing instability, spiral wave dynamics, bistability, spot replication~\cite{maini1997spatial,kondo2010reaction,Murray2002,KeenerSneyd1998} (and the references therein) by providing insights into mathematical mechanisms (e.g., linear, nonlinear, absolute, and convective instabilities) that give rise to pattern selection~\cite{cross1993pattern,borckmans2002diffusive,Pismen2006}. Among the more popular and exploited reactions are Belousov-Zhabotinsky and chlorite-iodide-malonic-acid (CIMA)~\cite{Epstein,borckmans2002diffusive}. Besides interests in chemical controls~\cite{CIreaction,borckmans2002diffusive,vanag2008design,szalai2012chemical}, these reactions are also used as phenomenological models for biological and ecological systems, examples of which include morphogenesis, cardiac arrhythmia, and vegetation in semi-arid regions~\cite{cross1993pattern,volpert2009reaction,Epstein,MeronEco}. 

An intriguing type of pattern formation phenomenon, demonstrating stationary spatial localization embedded in an oscillatory background, has been found experimentally in the CIMA reaction~\cite{kepper}. Such localized states have been observed in one- and two-space dimensions (1D and 2D, respectively)~\cite{borckmans1995localized,dewel1995pattern}, and attributed to a Turing core emerging in a Hopf background oscillating with a phase shift of $\pi$, a behavior that is typical in the vicinity of a codimension-2 bifurcation~\cite{kepper,FlipFlopand2Dspirals,Dual-mode}, a.k.a. a \textit{flip-flop} behavior or ``1D-\textit{spiral}''~\cite{perraud1993one,one-dimensional,Wit-flipflop}. The 2D localization was attributed to the phase singularity that forces a vanishing Hopf amplitude and thus in turn emergence of a Turing state~\cite{kepper,FlipFlopand2Dspirals,Dual-mode}. In the mathematical context, it was shown that the spatial localization in the 1D Hopf-Turing bifurcation~\cite{Brusselator,FlipFlopand2Dspirals} bears a similarity to the spatial localization mechanism in systems with a Turing-type (finite wavenumber) instability due to the \textit{homoclinic snaking} structure~\cite{Brusselator}. 

\begin{figure*}[tp]
	\begin{tabular}{ccc}
		\includegraphics[width=0.21\textwidth]{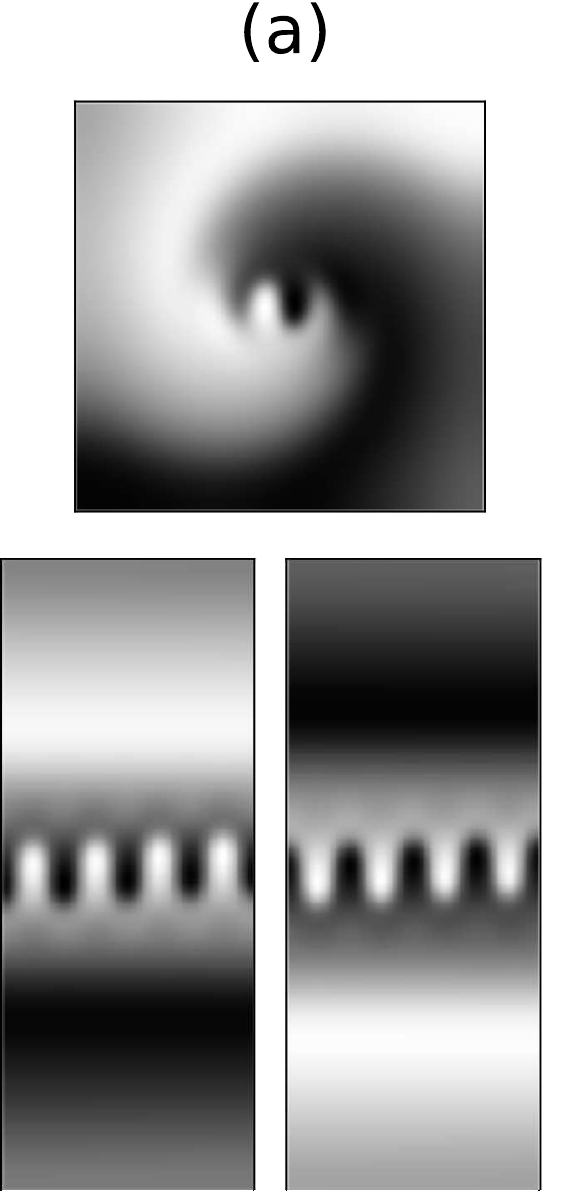}
		&\quad \includegraphics[width=0.3\textwidth]{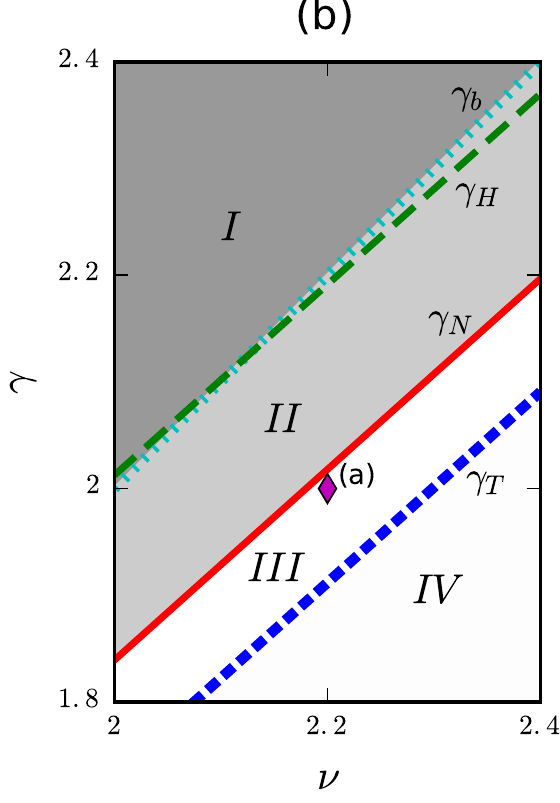}
		&\quad	\includegraphics[width=0.305\textwidth]{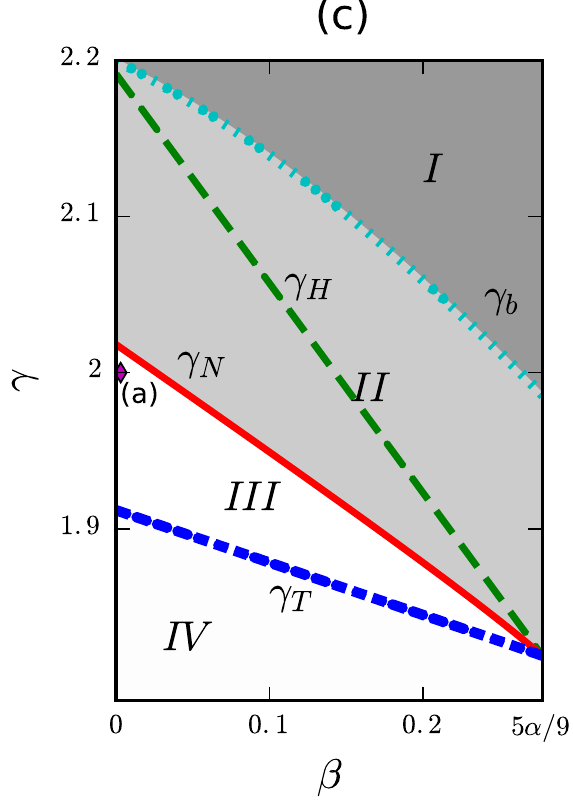}
	\end{tabular}
	\caption{(a) Direct numerical integration of~\eqref{FCGL} in 2D, showing snapshots of comb-like localizations inside a spiral wave (top panel) and in between $\pi$ shifted planar Hopf oscillations (bottom panel); light/dark colors indicate max/min value of the $Re(A)$ field, respectively. No-flux boundary conditions were used on spatial domains of $x\in [0,50]$, $y \in [0,50]$ (top panel) and $x\in [0,30]$, $y \in [0,75]$ (bottom panel).	(b) Applied forcing parameter plane ($\nu$-$\gamma$) showing four distinct regions (see text for details and definitions): Region I, which is the classical 2:1 frequency locking region, $\gamma>\gamma_b$; Region II, in which frequency locking is extended due to dominance of Turing states, $\gamma_N\leq\gamma\leq\gamma_b$; Region III, in which Hopf states are dominant but localized comb-like Turing states can emerge (solutions such as presented in (a), with location indicated by ({\large{\color{Purple}$\blacklozenge$}}) symbol), $\gamma_T\leq\gamma\leq\gamma_N$; Region IV that supports only unlocked oscillations, $\gamma<\gamma_T$. (c) Parameter plane ($\beta$-$\gamma$) showing the limit ($\beta_c=5\alpha/9$) of region III that corresponds to presence of pure Turing and Hopf modes (see text for details) while other symbols as in (b). Parameters: $\mu=0.5$, $\alpha=0.5$ and for (a) $\gamma=2$, $\nu=2.2$, $\beta=0$.}
	\label{Fig-multiple}
\end{figure*}

In this study, we focus on spatially localized comb-like structures in 2D, see for example Fig.~\ref{Fig-multiple}(a). We show that these localized states emerge via an alternative pinning mechanism over a much wider range of parameters and specifically, in a region where 1D homoclinic snaking is absent. We exploit the context of frequency locking and show that the Hopf-Turing localization in 2D further extends the resonant behavior outside of the resonance tongue~\cite{ArikLabyrinthine,ArikFCGL}. The paper is organized as follows: in the rest of the Introduction section, we briefly discuss the phenomenology of spatially extended Hopf-Turing patterns and their impact on the 2:1 resonance outside the {classical} locking region; in section~\ref{FlipFlopDynamics}, we overview the 1D flip-flop localization and show numerically that planar 2D comb-like localized states exist in a wide parameter range where flip-flop behavior is not present; in section~\ref{Comb-like-states}, we provide an alternative mechanism for localized comb-like states in 2D by deriving and analyzing three amplitude equations that represent a Hopf mode and two perpendicular Turing modes; then in section~\ref{spirals}, we exploit these insights to explain the comb-like spiral core; finally, we conclude in section~\ref{sec:concl}. 

\subsection{Coexistence of periodic stationary and temporal patterns: Hopf-Turing bifurcation}

{Consider a general reaction-diffusion type system:
	\begin{equation}\label{eq:genPDE}
		{\dfrac{{{\partial} \vec{u} }}{{{\partial}t}}}= {f \left(\vec{u} \right)}+\mathbf{D} \nabla^2 \vec{u},
	\end{equation}
	where $\vec{u}\equiv (u_1,\dots,u_N)$ are chemical subsets (with $N$ being an integer), $f(\vec{u})$ are functions containing linear and nonlinear terms that correspond to chemical reactions or interactions, and $\mathbf{D}$ is a matrix associated with diffusion and cross-diffusion~\cite{vanag2009cross}.}

In vicinity of the codimension-2 Hopf-Turing instability, the solution $\vec u(x,t)$ can be approximated by:
\[
\vec u \approx \vec u_\ast + \vec e_H H(\sqrt{\eps}x,\eps t)e^{i\omega_c t} + \vec e_T T(\sqrt{\eps}x,\eps t)e^{ik_c x}  + c.c.,
\]
where $\vec u_\ast$ is a spatially uniform state that goes through an instability, $c.c.$ is complex conjugate, $H$ and $T$ are slowly varying Hopf and Turing amplitudes in space and time, $\vec e_H$ and $\vec e_T$ are eigenvectors of the critical Hopf frequency ($\omega_c$) and Turing {wavenumber} ($k_c$) at a codimension-2 onset, respectively. Multiple time scale analysis using the above ansatz, leads to a generic set of Hopf and Turing amplitude equations~\cite{keener1976secondary,kidachi1980mode,Bose,de1999spatial}:
\begin{subequations}\label{eq:HT1D}
	\begin{eqnarray}
		\label{TXeqAH} \partiald{H}{t} & = & m_1 H 
		- m_2 |H|^2H - m_3 |T|^2 H + m_4 \doublepartiald{H}{x},\quad \quad\\
		\label{TXeqAX}\partiald{T}{t} & = & n_1 T 
		- n_2|T|^2 T - n_3 |H|^2T + n_4 \doublepartiald{T}{x},
	\end{eqnarray}
\end{subequations}
where $m_{1,2,3,4} \in \mathbb{C}$ and $n_{1,2,3,4} \in \mathbb{R}$.
Notably, system~\eqref{eq:HT1D} is a 1D reduction of~\eqref{eq:genPDE} and reproduces well the flip-flop behavior~\cite{kepper,FlipFlopand2Dspirals,one-dimensional,Wit-flipflop,dewel1995pattern}, i.e., a spatially localized Turing state embedded in $\pi$ shifted Hopf oscillations.  

\subsection{Frequency locking outside the resonant region}

The intriguing and rich dynamics of the Hopf-Turing bifurcation has been demonstrated not only in the CIMA reaction~\cite{kepper,CIreaction} but also has been found to be fundamental in broadening the 2:1 frequency locking behavior in the periodically forced Belousov-Zhabotinsky chemical reaction~\cite{ArikLabyrinthine}. Consequently, we focus here on the framework of frequency locking in spatially extended oscillatory media to study both the 2D Hopf-Turing localization and its relation to further increase of the 2:1 resonance region, in general.

Let us assume that system~\eqref{eq:genPDE} goes though a primary oscillatory Hopf instability and is also externally forced at a certain frequency~\cite{lin2004resonance}. Near the onset and depending on the forcing amplitude and frequency, the medium will exhibit either unlocked or locked oscillations which will obey the forced complex Ginzburg--Landau (FCGL) equation~\cite{gambaudo1985perturbation,elphick1987normal,CoulletEmilsson1992,elphick1999multiphase}:
\begin{equation}\label{FCGL}
	\partiald{A}{t} = (\mu+i\nu)A - (1+ i\beta)|A|^2A + \gamma {\bar A}^{n-1} + (1+i\alpha)\nabla^2A,
\end{equation}
where $A$ is a weakly varying in space and time complex amplitude of the primary Hopf mode, $\bar A$ complex conjugate, $n$ is an integer associated with the $n:1$ resonance, $\mu$ is the distance from the Hopf onset, $\nu$ is the difference between natural and the forcing frequencies, and $\gamma$ is the forcing amplitude. {In this context, frequency locking corresponds to asymptotically stationary solutions to~\eqref{FCGL}. Since the Hopf-Turing bifurcation in~\eqref{FCGL} arises only for $n=2$, the study of Hopf-Turing spatially localized states applies to only in 2:1 resonant case.}

In the 2:1 resonance case, Eq.~\ref{FCGL} admits two uniform non-trivial ($\pi$ shifted) solutions that exist for~\cite{CoulletEmilsson1992,ArikLabyrinthine,ArikFCGL}
\begin{equation}
	\gamma>\gamma_b=\frac{|\nu -\mu\beta|}{\sqrt{1-\beta^2}}.
\end{equation}
This bistability region is commonly called the classical (Arnol'd) 2:1 resonance tongue even in the context of the spatially extended media, see region $I$ in Figs.~\ref{Fig-multiple}(b,c). {Moreover, bistability of uniform $\pi$ shifted states, can also lead to formation of inhomogeneous solutions~\cite{coullet1990breaking,gomila2001stable,ArikFCGL,burke2008classification}, such as labyrinthine patterns, spiral waves, and spatially localized (a.k.a. oscillons).}
However, it has been shown that nonuniform 2:1 resonant patterns may in fact exist also {outside the 2:1 resonance~\cite{ArikLabyrinthine}, $\gamma<\gamma_b$, i.e., in a region where stationary non-trivial uniform solutions are absent}. The resonant condition {is obtained through stripe (Turing state) nucleation} due to the propagation of a Hopf-Turing front, i.e., an interface that bi-asymptotically connects Hopf and Turing states, {as shown in Fig.~\ref{Fig-multiple}(a)}. 

\begin{figure}[tp]
	\includegraphics[width=0.45\textwidth]{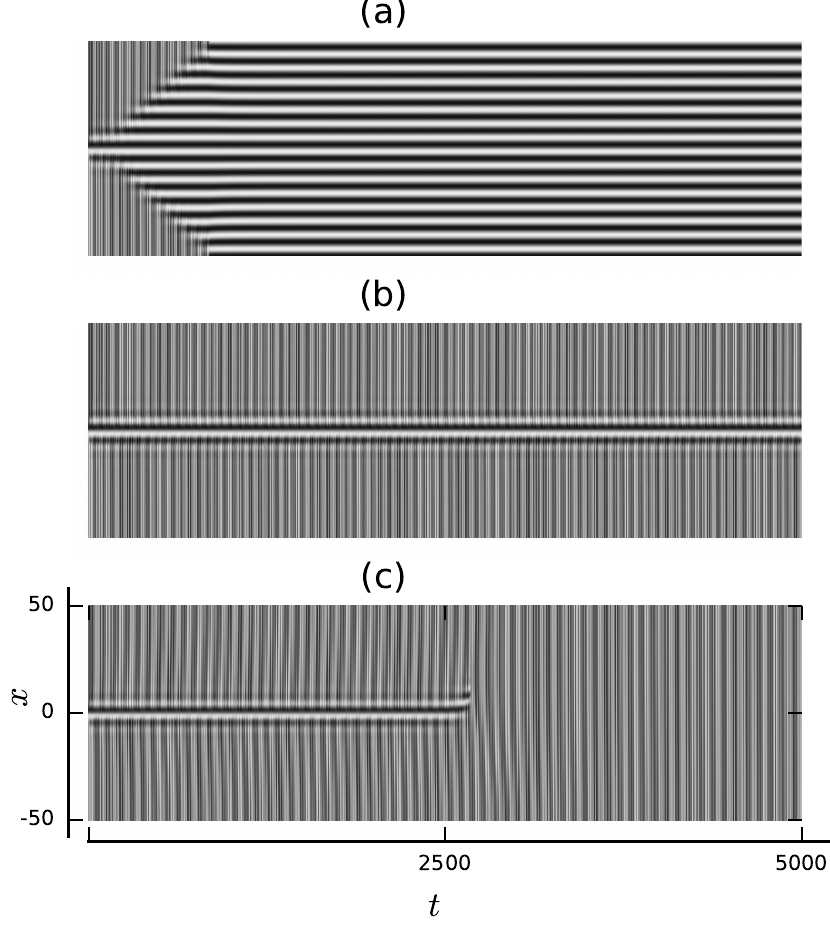}
	\caption{Direct numerical integration of \eqref{FCGL} in 1D, showing the flip-flop and the depinning dynamics; light/dark colors indicate max/min value of the  $Re(A)$ field, respectively. {(a) $\gamma=2.015>\gamma_N \simeq 2.01$ (region II in Fig.~\ref{Fig-multiple}), (b) $\gamma=\gamma_N \simeq 2.01$, (c) $\gamma=2.002<\gamma_N \simeq 2.01$ (region III in Fig.~\ref{Fig-multiple})}, while other parameters $\mu=0.5$, $\nu=2.2$, $\beta=0$, $\alpha=0.5$.}
	\label{Fig-FCGLflipflop}
\end{figure}

The codimension-2 Hopf-Turing bifurcation is an instability of the trivial uniform state $A=0$ at $\mu=0$ and $\gamma=\gamma_c$~\cite{ArikFCGL}, with $\omega_c =  \nu\alpha/\rho$, $k_c^2  =  \nu\alpha/\rho^2$, $\gamma_c  =  \nu/\rho$ and $\rho = \sqrt{1+\alpha^2}$. Notably, the Hopf-Turing bifurcation occurs outside the resonance region as $\gamma_c<\gamma_b$. Multiple time scale analysis resulted with coefficients~\cite{ArikFCGL} for the Hopf-Turing amplitude equations~\eqref{eq:HT1D}:
\begin{align*}
	m_1=&\mu - i\frac{\gamma -\gamma_c}{\alpha}, m_2= 4+i2\beta \frac{2\rho^2+1}{\alpha\rho}, \\
	m_3=& 8\rho(\alpha+\rho) \\ &+ i \{4\beta \frac{2\alpha\rho(\alpha + \rho) + 3\rho + \alpha}{\alpha} - 4(\alpha+\rho)\}, \\
	m_4=& 1+i\rho, n_1=\mu + \rho\frac{\gamma -\gamma_c}{\alpha}, \\
	n_2=& 6\rho(\alpha+\rho) \braa{1-\frac{\beta}{\alpha}}, 
	n_3= 4\braa{2-3\frac{\beta}{\alpha}}, n_4= 2\rho^2.
\end{align*}
Specifically, stability analysis of pure Hopf and Turing modes showed that for $\beta=\beta_c<5\alpha/9$ these two uniform states coexist and thus it is possible to form a heteroclinic connection between them~\cite{ArikFCGL}, i.e., a front solution. The Hopf-Turing front is stationary {(Fig.~\ref{Fig-FCGLflipflop}(b))} at 
\begin{equation}\label{eq:gammaN}
	\gamma_N = \gamma_c+ \frac{\mu}{\rho}\bigg[\sqrt{\frac{3}{4}(2\alpha-3\beta)(\alpha-\beta)}\,\,-\,\alpha\bigg],
\end{equation} 
and propagates otherwise~\cite{BodeFronts}, with $\gamma>\gamma_N$ the Turing state invades the Hopf state {(Fig.~\ref{Fig-FCGLflipflop}(a)) and vice-verse (Fig.~\ref{Fig-FCGLflipflop}(c)). Moreover,} the presence of the stationary front serves as an organizing center for the homoclinic snaking phenomena~\cite{Brusselator} that would be discussed in the next section.

{The dominance of the asymptotically stationary Turing mode in region II, $\gamma_N<\gamma<\gamma_b$, extends thus, the classical frequency locking domain (region I) once spatially extended patterns are formed~\cite{ArikLabyrinthine,yochelis2004frequency}. Notably, Turing type solutions are in fact standing-waves in the context of the original system~\eqref{eq:genPDE}. Figures~\ref{Fig-multiple}(b,c) show the classical resonance region for a single oscillator (region $I$) and the extended frequency locked region due to the dominance of a spatially extended Turing mode (region $II$). Our interest is thus, in the unlocked region III ($\gamma_T<\gamma<\gamma_N$ in Figures~\ref{Fig-multiple}(b,c)) where, despite the Hopf mode dominance (i.e., Hopf state is favorable over the Turing state), 2D resonant \textit{localized} comb-like states may still form (see Figure~\ref{Fig-multiple}(a)), with}
\begin{equation}
	\gamma_T  = \gamma_c-\frac{\mu}{4\rho}(\alpha+3\beta),
\end{equation}
which is the stability onset of the Turing mode~\cite{ArikFCGL}. For $\gamma<\gamma_T$ only Hopf oscillations persist, i.e., region $IV$.

In what follows, we use $\gamma$ as a control parameter while keeping all other parameters constant. Notably, we limit the scope to the coexistence region between the Hopf and Turing modes~\cite{ArikFCGL}, with ($\beta<5\alpha/9$) and $\gamma_T<\gamma<min\{\gamma_H,\gamma_b\}$, where $\gamma_H=\gamma_c+\mu(\alpha-3\beta)/\rho$. As such the localized 2D solutions in region $III$ are resonant states and thus extend further the frequency locking boundary, as portrayed in Figure~\ref{Fig-multiple}.

\section{Flip-flop dynamics and depinning} \label{FlipFlopDynamics}

Hopf-Turing spatial localization, pinning, and depinning in 1D have been studied in detail by Tzou \textit{et al.} (2013), who have shown the relation to the homoclinic snaking phenomenon~\cite{Brusselator}. Specifically, two snaking behaviors were outlined: 
\begin{description}
	\item[Standard (vertical) snaking] if the Turing mode is embedded in Hopf background that is oscillating in phase;
	\item[Collapsed snaking] if the Turing mode is embedded in Hopf background that oscillates with a phase shift of $\pi$. This case is also known as the "flip-flop" behavior.
\end{description}
Both cases form in the vicinity of a stationary front (Maxwell-type heteroclinic connection) between the Hopf (oscillatory) and the Turing (periodic) states, i.e., around $\gamma=\gamma_N$ in the context of FCGL [see Eq.~\ref{eq:gammaN}]. The width of the snaking regime is, however, rather narrow and depinning effects become dominant at small deviations from $\gamma_N$. Indeed, numerical integrations of~\eqref{FCGL} confirm this result also in the context of FCGL (see Fig.~\ref{Fig-FCGLflipflop}): for $\gamma>\gamma_N$ ($\gamma<\gamma_N$) the Turing (Hopf) state invades the Hopf (Turing) state~\cite{ArikFCGL}, and thus the localized Turing state (centered at $x=0$) expands (collapses), respectively.

\begin{figure*}[tp]
	\begin{tabular}{ccc}
		\includegraphics[width=0.19\textwidth]{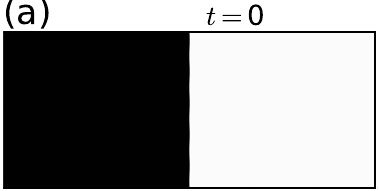}\hspace{1.3cm}
		&\includegraphics[width=0.19\textwidth]{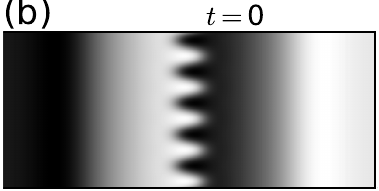}\hspace{0.8cm}
		&\quad	\includegraphics[width=0.37\textwidth]{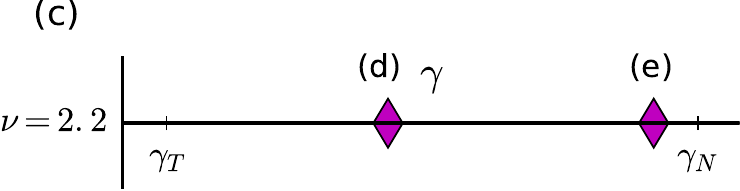}
	\end{tabular}
	\begin{tabular}{c}
		\vspace{0.1cm}\\
		\includegraphics[width=0.9\textwidth]{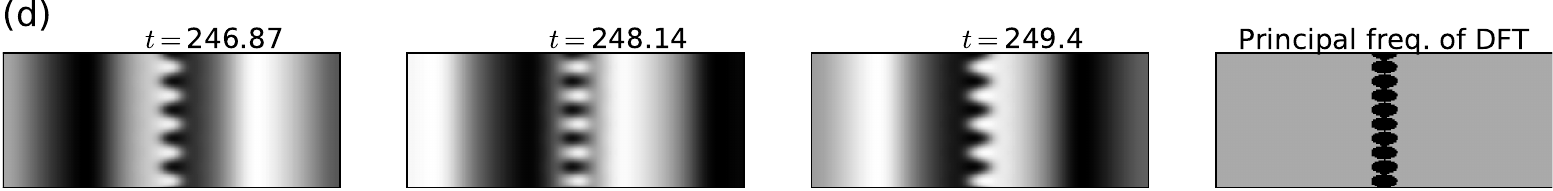}\\
		\vspace{0.1cm}\\
		\includegraphics[width=0.9\textwidth]{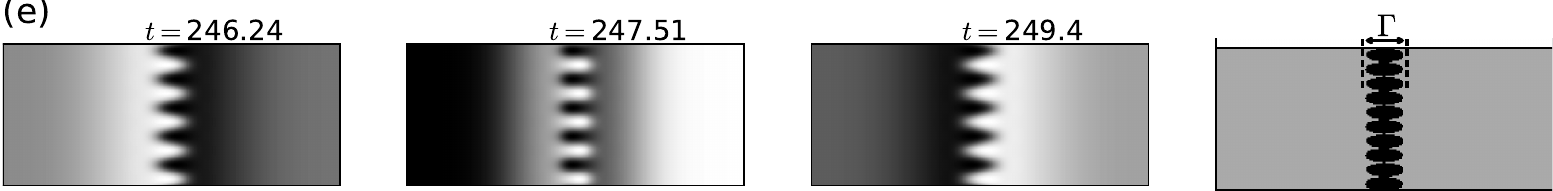}
	\end{tabular}
	\caption{Direct numerical integration of~\eqref{FCGL} in 2D using (a) and (b) as initial conditions (at $t=0$) for (d) $\gamma=1.94$ and (e) $\gamma=2$, respectively, as also indicated in top panel (c) by ({\large{\color{Purple}$\blacklozenge$}}) symbols, respectively. The $Re(A)$ field is presented where light/dark colors indicate max/min values, respectively. The far right column represents the frequencies with the highest amplitude obtained by discrete Fourier transform for the case (see text for details), respectively. Here, the gray color corresponds to $4/(\tau\Delta t)$, where $\tau=40$ is the number of elements in the evaluated time series and $\Delta t=0.633$ corresponds to time steps within the interval, thus, the time window length which was taken corresponds to $\tau(\Delta t)=25.32$. The width of the comb-like state is then approximated by $\Gamma$ which corresponds to the black color of vanishing amplitude value. The calculations were conducted on a spatial domain $x\in [0,75]$, $y \in [0,31.5]$, with no-flux boundary conditions in $x$ and periodic in $y$. Parameters: $\mu=0.5$. $\nu=2.2$, $\beta=0$, $\alpha=0.5$.}
	\label{front}
\end{figure*}

Onthe other hand, the robustness of comb-like structures (e.g., in spiral waves) as compared to the narrow existence in the parameter space of the flip-flop, suggests that the emergence mechanism is distinct. Indeed, direct numerical simulations in 2D show that comb-like localized patterns emerge in a parameter range in which flip-flop does not coexist $\gamma_T<\gamma < \widetilde \gamma_N$, where $\widetilde \gamma_N \lessapprox \gamma_N$ is considered to be the left limit of the depinning region and computed here numerically. Notably, since the snaking region is very narrow as compared to the rest of the domain, we define in what follows region $III$ to lie within $\gamma_T<\gamma<\gamma_N$. Moreover, the width ($\Gamma$) of the localized comb-like region increases with $\gamma$, as shown in Fig.~\ref{front}. To quantify $\Gamma$, we employed a discrete Fourier transform (DFT) for every grid point. The dark shading marks the frequency with the highest contribution, as shown in the furthermost right panels in Fig. \ref{front}. These results indeed confirm that comb-like states (see Fig.~\ref{Fig-multiple}) are related to a distinct 2D pinning mechanism and not just a spatial extension of flip-flop dynamics~\cite{perraud1993one,one-dimensional,Wit-flipflop,FlipFlopand2Dspirals,Dual-mode}. 


\section{Comb-like localized states}\label{Comb-like-states}

In this section, we show that localized comb-like states are formed due to pinning of a Turing mode that is perpendicular to the $\pi$ phase shifted Hopf oscillations. Namely, we look for planar localized states, as shown in Figs.~\ref{front}(d,e). At first, we derive the respective amplitude equations, then we obtain uniform solutions along with their stability properties, and finally confirm the results by direct numerical integrations.

\subsection{Weakly nonlinear analysis}\label{HopfTuringequations}

Derivation of amplitude equations in 2D follows, in fact the same steps as for the 1D case but with two Turing complex amplitudes (both varying slowly in space and time) 
\begin{eqnarray}\label{eq:3ampl_anz}
	\nonumber \colvec{Re(A)}{Im(A)} &\approx & \colvec{(1+i\alpha)/\rho}{1} H\braa{\sqrt{\eps}x,\sqrt{\eps}y,\eps t}e^{i\omega_c t} \\
	&& + \colvec{(\alpha+\rho)}{1}T_\parallel\braa{\sqrt{\eps}x,\sqrt{\eps}y,\eps t}e^{ik_c x} \\
	\nonumber  && +  \colvec{(\alpha+\rho)}{1} T_\perp\braa{\sqrt{\eps}x,\sqrt{\eps}y,\eps t}e^{ik_c y} \\
	\nonumber  && + c.c. + h.o.t.,
\end{eqnarray}
where $h.o.t.$ stands for high order terms. The two Turing modes $T_\parallel$ and $T_\perp$ are defined as parallel and perpendicular to the considered $\pi$ phase shifted Hopf oscillations (hereafter, Hopf front), respectively. Following the multiple time scale method (see~\cite{ArikFCGL} for details), we obtain (after some algebra that is not shown here)
\begin{subequations}\label{eq:3ampl}
	\begin{eqnarray}
		\nonumber	\label{TXeqAH} \partiald{H}{t} & = & m_1 H - m_2 |H|^2H - m_3 \braa{|T_\perp|^2+|T_\parallel|^2} H \\
		&& + m_4 \nabla^2 H, \\
		\nonumber	\label{TXeqAX}\partiald{T_\parallel}{t} & = & n_1 T_\parallel - n_2\braa{|T_\parallel|^2+2|T_\perp|^2} T_\parallel - n_3 |H|^2T_\parallel \\
		&&  + n_4 \doublepartiald{T_\parallel}{x},\\
		\nonumber	\label{TXeqAY}\partiald{T_\perp}{t} & = & n_1 T_\perp - n_2\braa{2|T_\parallel|^2+|T_\perp|^2} T_\perp - n_3 |H|^2T_\perp \\
		&& +n_4 \doublepartiald{T_\perp}{y}.
	\end{eqnarray}
\end{subequations}

Besides the standard Hopf-Turing solutions~\cite{ArikFCGL}, uniform solutions to~\eqref{eq:3ampl} that involve non-vanishing $T_\perp$ contributions, are obtained through the amplitudes $(|H|,|T_\parallel|,|T_\perp|):=(R_H,R_\parallel,R_\perp)=(\widetilde R_H,\widetilde R_\parallel,\widetilde R_\perp)$:

\begin{itemize}
	\item Pure Turing modes (stripes),
	\begin{equation}\label{eq:pureT}
		\widetilde R_{\perp} = \sqrt{\dfrac{\mu \alpha + \rho (\gamma-\gamma_c)}{6\rho(\alpha+\rho)(\alpha-\beta)}}, \quad \widetilde R_H =\widetilde R_{\parallel}= 0;
	\end{equation}
	
	\item Unstable mixed Turing mode (stationary squares),
	\begin{equation}\label{eq:MT}
		\widetilde R_\perp=\widetilde R_\parallel= \sqrt{\dfrac{\mu \alpha + \rho (\gamma-\gamma_c)}{18\rho(\alpha+\rho)(\alpha-\beta)}}, \quad \widetilde R_H = 0;
	\end{equation}
	
	\item Unstable mixed Hopf-Turing mode (oscillating squares),
	\begin{eqnarray}\label{eq:MHT}
		\nonumber	\widetilde R_H &=& \frac{1}{2}\sqrt{\dfrac{(18\beta-2\alpha)\mu + 16\rho (\gamma-\gamma_c)}{14\alpha-30\beta}}, \\
		\widetilde R_{\perp}=\widetilde R_{\parallel} &=& \sqrt{\dfrac{(\alpha-3\beta)\mu -\rho (\gamma-\gamma_c)}{\rho(\alpha+\rho)(14\alpha-30\beta)}}.
	\end{eqnarray}
\end{itemize}
\begin{figure*}[tp]
	\includegraphics[width=0.9\textwidth]{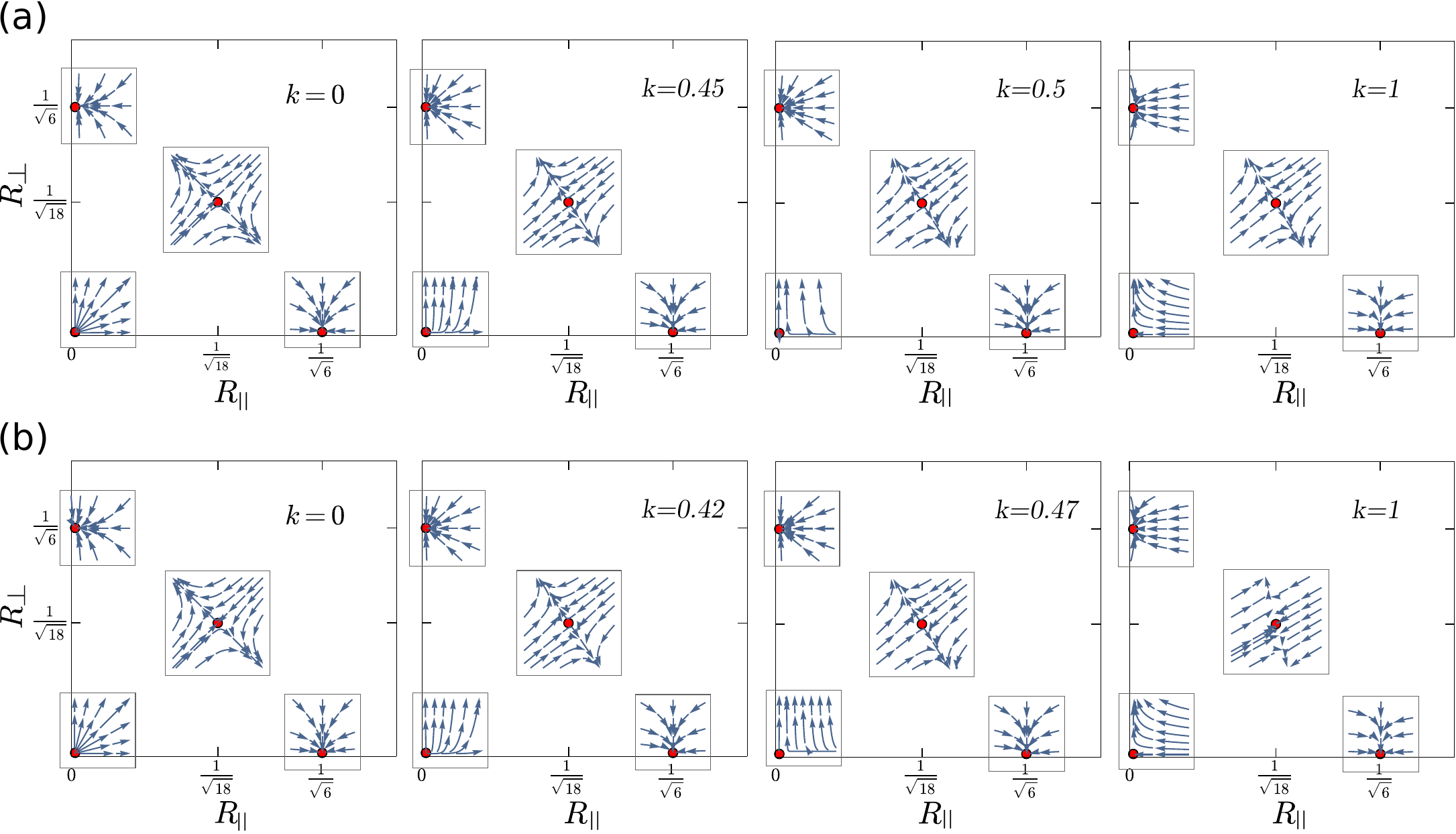}
	\caption{Streamline of linearized vector flow for~\eqref{!HTu} about four fixed points at (a) $\gamma=1.94$ and (b) $\gamma=2$, projected on the $(R_\parallel,R_\perp)$ plane, and calculated for four distinct wavenumbers, as shown from left to right: $k \to 0$, $k\lesssim k_f$, $k\gtrsim k_f$, and $k > k_f$, where $k_f$ is given in~\eqref{eq:k_f}. The axis units are $\sqrt{6}\widetilde{R}_{\parallel,\perp}$ according to~\eqref{eq:pureT}, and parameters: $\mu=0.5$, $\nu=2.2$, $\beta=0$, $\alpha=0.5$.}
	\label{diagram}
\end{figure*}

\subsection{Stability and Hopf fronts}

{After obtaining uniform solutions, we proceed to the selection mechanism by focusing on the spatial symmetry breaking that is induced by the Hopf front.} Consequently, we associate~\eqref{eq:3ampl} with only one spatial dependence (here we use $x$), which corresponds to the direction of a Hopf front. In addition, for convenience we use polar form $H=R_H\exp{(i\Phi)}$ and consider only the amplitudes of Turing fields (due to spatial dependence $H$ cannot be decoupled as for $T_{\parallel,\perp}$). Hence, system~\eqref{eq:3ampl} becomes
\begin{subequations}\label{!HTu}
	\begin{eqnarray}
		\label{HTRH}
		\nonumber \partiald{R_H}{t}  & = &  \mu R_H -4 R_H^3 - 8\rho\braa{\alpha+\rho}\braa{R_{\parallel}^2 + R_{\perp}^2}R_H\\
		\nonumber &&-\Bra{\braa{\partiald{\Phi}{x}}^2+\rho \doublepartiald{\Phi}{x}}R_H - 2\partiald{\Phi}{x}\partiald{R_H}{x} ,\\
		&&+ \doublepartiald{R_H}{x} \\
		\label{HTArgH}
		\nonumber \partiald{\Phi}{t} & = & -\frac{\gamma -\gamma_c}{\alpha}- \nu_1R_H^2-\nu_2\braa{R_{\parallel}^2 + R_{\perp}^2} \\
		\nonumber &&  -\Bra{\rho\braa{\partiald{\Phi}{x}}^2-  \doublepartiald{\Phi}{x}} + \frac{2}{R_H}\partiald{\Phi}{x}\partiald{R_H}{x}\\
		&& + \frac{\rho}{R_H} \doublepartiald{R_H}{x},   \\
		\label{HTrx}
		\nonumber \partiald{R_\parallel}{t} & = & \Bra{\mu + \frac{\rho(\gamma-\gamma_c)}{\alpha}}R_\parallel - 4\bigg(2-3\frac{\beta}{\alpha}\bigg) R_H^2R_\parallel\\
		\nonumber && - 6\rho(\alpha+\rho)\bigg(1-\frac{\beta}{\alpha}\bigg)\braa{R_\parallel^2 + 2R_\perp^2}R_\parallel \\
		&& + 2\rho^2\doublepartiald{R_\parallel}{x}, \\
		\label{HTry}  
		\nonumber \partiald{R_\perp}{t} & = & \Bra{\mu + \frac{\rho(\gamma-\gamma_c)}{\alpha}}R_\perp - 4\bigg(2-3\frac{\beta}{\alpha}\bigg) R_H^2R_\perp \\
		&& - 6\rho(\alpha+\rho)\bigg(1-\frac{\beta}{\alpha}\bigg)\braa{2R_\parallel^2 + R_\perp^2}R_\perp, 
	\end{eqnarray}
\end{subequations}
where $\nu_1={2\beta}(2\rho^2+1)/(\alpha \rho)$, and $\nu_2 = {4\beta}[2\alpha\rho(\alpha + \rho) + 3\rho + \alpha]/{\alpha} - 4(\alpha+\rho)$. Notedly, the spatial symmetry breaking is reflected in~\eqref{!HTu} through the absence of a diffusive term.
\begin{figure*}[tp]
	\begin{tabular}{c}
		\includegraphics[width=0.9\textwidth]{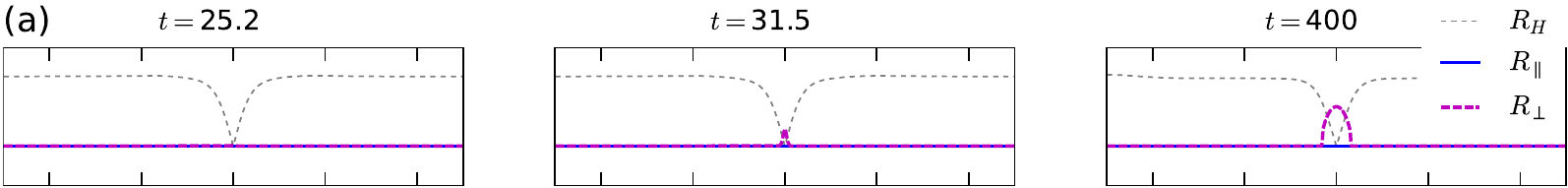}\\
		\vspace{0.02cm}\\
		\includegraphics[width=0.9\textwidth]{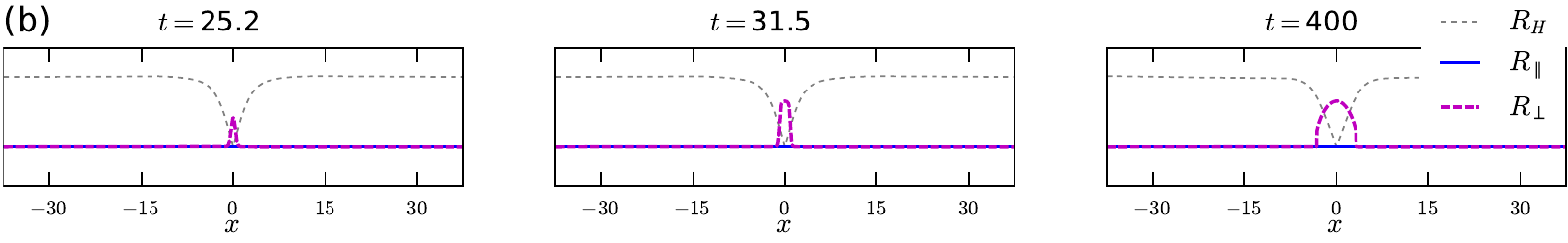}
	\end{tabular}
	\caption{Direct numerical integration of~\eqref{eq:3ampl} showing snapshots of the amplitudes $(R_H,R_\parallel,R_\perp)$ at (a) $\gamma=1.94$ and (b) $\gamma=2$ (as also indicated in Fig.~\ref{front}(c)). The calculations were conducted on a spatial domain $x\in [0,75]$, under no-flux boundary conditions, and parameters: $\mu=0.5$. $\nu=2.2$, $\beta=0$, $\alpha=0.5$.}
	\label{HTfrontTy}
\end{figure*}
\begin{figure}[tp]
	\includegraphics[width=0.4\textwidth]{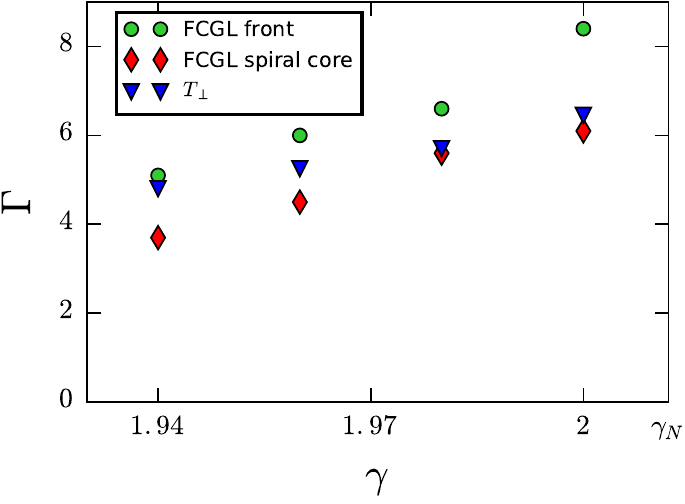}
	\caption{{Width ($\Gamma$) of the comb-like state as obtained from direct numerical integrations of~\eqref{FCGL} in two space dimensions, and a localized $T_\perp$ Turing state using~\eqref{eq:3ampl}. Parameters: $\mu=0.5$. $\nu=2.2$, $\beta=0$, $\alpha=0.5$. Notably, although $\mu$ is about order 1 from the co-dimension 2 onset ($\mu=0$), the width of $T_\perp$ obtained from integration of~\eqref{eq:3ampl} is well within the range of comb-like solutions to~\eqref{FCGL}.}}
	\label{Fig-yTwidth}
\end{figure}

{The fixed point analysis (linear stability to spatially uniform perturbations)} of~\eqref{!HTu} is identical to the 1D case~\cite{ArikFCGL}, and thus, {the pure Hopf and pure Turing} coexistence regime remains the same, $\beta<5/(9 \alpha)$. However, to gain insights into the emergence of the $T_\perp$ mode at the Hopf front region, we examine the linear stability of the trivial solution $({\widetilde{R}_H},{\widetilde{R}_\parallel},{\widetilde{R}_\perp})$=$(0,0,0)$ to non-uniform perturbations, for which the Hopf amplitude and phase can be decoupled:
\begin{equation}\label{eq:disp}
	\left( {\begin{array}{*{20}c}
			{R_H}  \\
			{R_\parallel}  \\
			{R_\perp}  \\
		\end{array}} \right) -
		\left( {\begin{array}{*{20}c}
				{\widetilde{R}_H}  \\
				{\widetilde{R}_\parallel}  \\
				{\widetilde{R}_\perp}  \\
			\end{array}} \right) \propto e^{\sigma t -ikx}+c.c.,
		\end{equation}
		where $\sigma$ is a growth rate of respective wavenumbers, $k$. Substitution of~\eqref{eq:disp} in~\eqref{!HTu} and solving to a leading order, yields three dispersion relations:
		\begin{eqnarray}
			\sigma_H&=&\mu-k^2,\\
			\sigma_0&=&\mu +\rho \dfrac{\gamma-\gamma_c}{\alpha}.\\
			\sigma_k&=&\mu +\rho \dfrac{\gamma-\gamma_c}{\alpha}-2\rho^2k^2.
		\end{eqnarray}
		
		As expected, the three growth rates show instability for $k=0$, where the vector flow is rather isotropic (Fig.~\ref{diagram}). For completeness, we have computed the trajectories after linearizing~\eqref{!HTu} about all the fixed points involving $\widetilde{R}_\perp$, and show the projection on the $(R_\parallel,R_\perp)$ plane. The results are consistent with the stability of pure Turing modes~\eqref{eq:pureT} and the saddle for a mixed Turing mode~\eqref{eq:MT}.
		As $k$ is increased, we observe a symmetry breaking between $\sigma_0$ and $\sigma_k$, which occurs for $\sigma_k=0$ or equivalently for
		\begin{equation}\label{eq:k_f}
			k^2_f = \frac{ \mu\alpha + \gamma\rho - \nu}{ 2\rho^2}.
		\end{equation}
		Figure~\ref{diagram} shows that the perturbations about $(R_\parallel,R_\perp)=(0,0)$ favor the attractor $(R_\parallel,R_\perp)=(0,{\widetilde R}_\perp)$. The increasing value of the wavenumber is also consistent with the basin of attraction which corresponds to a rather narrow spatial region due to the Hopf front location, i.e., already for $k=1$ the flow indicates ultimate preference toward $(R_\parallel,R_\perp)=(0,{\widetilde R}_\perp)$. 
		\begin{figure*}[htb]
			\begin{tabular}{c}
				\includegraphics[width=0.9\textwidth]{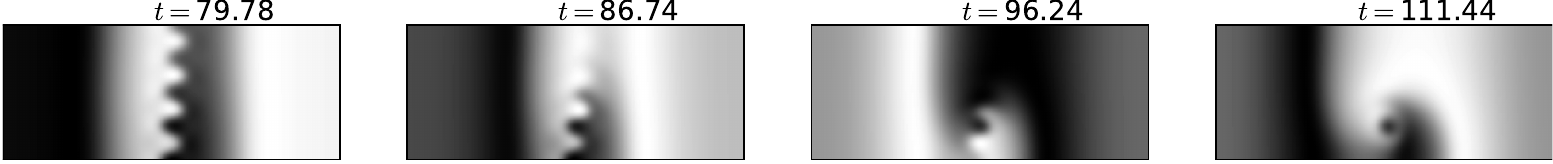}
			\end{tabular}
			\caption{Direct numerical integration of~\eqref{FCGL} in 2D, showing snapshots of the $Re(A)$ field for $\gamma=1.94<\gamma_N$; light/dark colors indicate max/min values, respectively. Initial condition was the same illustrated in Fig.~\ref{front}(a). The calculations were conducted on a spatial domain $x\in [0,75]$, $y \in [0,30]$ with no-flux boundary conditions. Other parameters: $\mu=0.5$. $\nu=2.2$, $\beta=0$, $\alpha=0.5$.}
			\label{FrontSpiral}
		\end{figure*}
		\begin{figure*}[tp]
			\begin{tabular}{c}
				\includegraphics[width=0.9\textwidth]{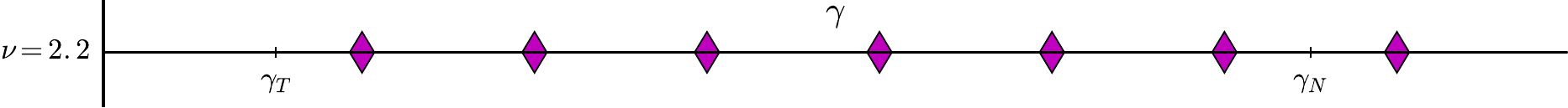}\\
				\vspace{0.1cm}\\
				\includegraphics[width=0.9\textwidth]{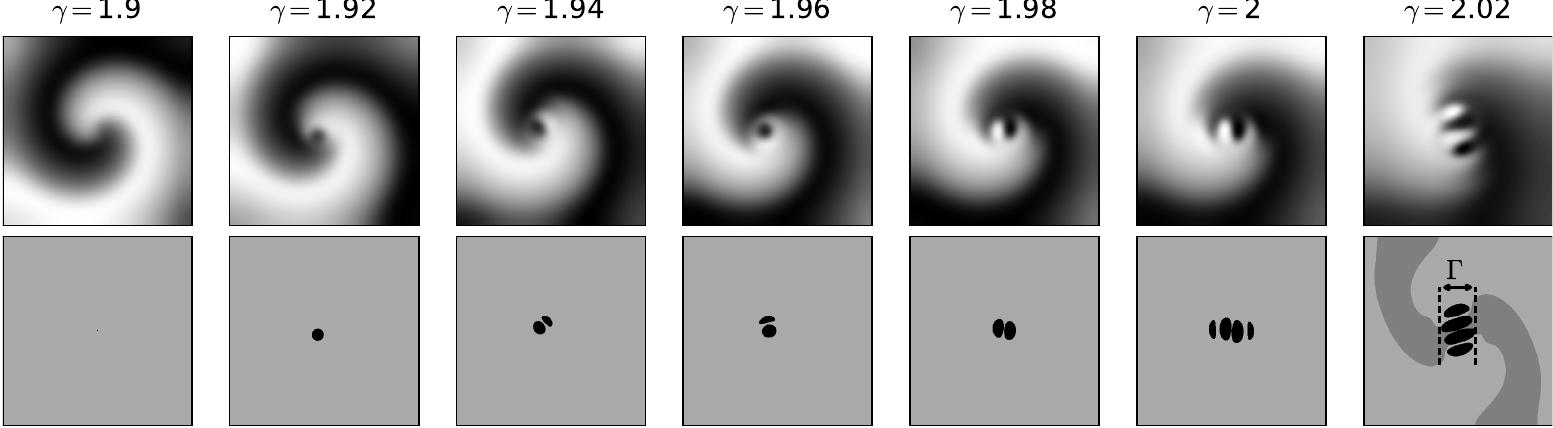}
			\end{tabular}
			\caption{Direct numerical integration of~\eqref{FCGL} in 2D, showing snapshots of the $Re(A)$ field for various values of $\gamma$ as indicated for each frame in the middle panel and also indicated in top panel by ({\large{\color{Purple}$\blacklozenge$}}) symbols, respectively; light/dark colors indicate max/min values, respectively. Bottom panel represents the frequencies with highest amplitude obtained by discrete Fourier transform for the case in the middle panel (see text and Fig.~\ref{front} for details), respectively. The calculations were conducted on a spatial domain $x\in [0,50]$, $y \in [0,50]$ with no-flux boundary conditions. Parameters: $\mu=0.5$. $\nu=2.2$, $\beta=0$, $\alpha=0.5$.}
			\label{spiralsDFT}
		\end{figure*}
		
		\subsection{Numerical results}
		
		Next, we check the above obtained results vs. direct numerical integrations of~\eqref{eq:3ampl}. We set a sharp Hopf front by using the Hopf amplitude as an initial condition: $R_H(x=0 \ldots \pm L)=\mp \widetilde{R}_H$ and $R_\parallel = R_\perp = 0$. Due to diffusion in the Hopf field, at first a front solution is indeed formed and after additional transient an asymptotic localized $T_\perp$ Turing state emerges inside the Hopf front region, as shown in Fig.~\ref{HTfrontTy}. The results are in accord with the linear stability analysis of the trivial state, showing that perturbations at the mid front location $x=0$, are in the basin of attraction of the fix point $(R_H, R_\parallel, R_\perp)=(0,0,\widetilde R_\perp)$ which corresponds to the comb-like structure in Fig.~\ref{front}. The width of the $T_\perp$ Turing localized state increases with $\gamma$, which agrees well with numerical integration of~\eqref{FCGL}, as shown in Fig.~\ref{Fig-yTwidth}. This could be an important feature that explains the different size of Turing core embedded in spiral waves. We note that in the narrow vicinity of $\gamma_N$ both Turing modes, $T_\perp$ and $T_\parallel$, coexist and can emerge depending on the initial perturbations within the Hopf front.
		
		{The Turing mode that is parallel to the Hopf front ($T_\parallel$) is being described by a partial differential equation~\eqref{HTrx} and thus in the absence of a pinning mechanism such as, homoclinic snaking, is being directly subjected to diffusive fluxes, which are overtaken by the oscillatory Hopf mode. On the other hand, the perpendicular Turing mode ($T_\perp$) obeys only a local dynamics via an ordinary differential equation~\eqref{HTry}. The spatial decoupling in~\eqref{HTry} allows thus, under certain initial conditions, pinning of the Hopf amplitude (local selection of the $(R_H, R_\parallel, R_\perp)=(0,0,\widetilde R_\perp)$ fixed point), which in turn results effectively in a $\pi$ phase shifted front.} This however, is highly sensitive to domain size or initial conditions due to the secondary zig-zag and Eckhaus instabilities~\cite{cross1993pattern,MeronEco}. For example, the length of $y$ dimension should be an integer of the typical wavenumber which is close to $k_c$ and within the Eckhaus stable regime; details of the Busse balloon for this problem are given in~\cite{ArikFCGL}. If this condition is not fulfilled, the nonlinear terms become dominant and the comb-like structures are destroyed and instead a spiral wave with a comb-like core is formed, see Fig.~\ref{FrontSpiral}.  
		
		\section{Spiral waves with comb-like core}\label{spirals}
		
		To capture the emergence of the Turing core embedded inside a Hopf spiral, we start with a pure Hopf spiral wave obtained for $\gamma=1.9$, as an initial condition. As for the planar front case (Fig.~\ref{front}), direct numerical integration of~\eqref{FCGL} for $\gamma_T<\gamma<\gamma_N$ shows formation of a Turing spot inside the core due to the vanishing amplitude of the Hopf amplitude within that region, see Fig.~\ref{spiralsDFT}. As expected, also here, the Turing spot size increases with $\gamma$. 
		
		To quantify the size of the Turing core, we use again DFT for each grid point within a window of 128 time steps, where each time step is made out of 100 discrete time iterations. Consequently, each grid point corresponds to a 128-dimensional vector with the amplitude calculated from DFT, where only the elements with the highest amplitude value are selected, as shown in the bottom panel of Fig.~\ref{spiralsDFT}. The frequency contrast allows us to define a criterion for the width ($\Gamma$) of the Turing spot. The spiral core width is found to agree with results obtained via the planar front initial condition, as shown Fig.~\ref{Fig-yTwidth}. In the depinning region above the stationary Hopf-Turing front condition $\gamma_N<\gamma<\gamma_b$, the spiral core expands by invasion into the Hopf oscillations due to the dominance of the Turing mode and the domain is filled with a periodic pattern~\cite{ArikLabyrinthine} (not shown here).
		
		\section{Conclusions}\label{sec:concl}
		
		In summary, we have presented a distinct pinning mechanism for 2D spatial localization that is associated with the emergence of {\emph{comb-like}} structures embedded in a temporally oscillatory background. These spatially localized states emerge in a planar form inside $\pi$ phase shift oscillations (Fig.~\ref{front}) or as a spiral wave core (Fig.~\ref{spiralsDFT}). The mechanism requires coexistence of periodic stripes in both $x$ and $y$ directions and uniform oscillations, a behavior that is typical in the vicinity of a codimension-2 Hopf-Turing bifurcation. Unlike the homoclinic snaking mechanism that gives rise to localized states over a narrow range of parameters about a stationary Hopf-Turing front in 1D (a.k.a. flip-flop dynamics)~\cite{Brusselator}, the comb-like states are robust (i.e., do not require any Maxwell type construction) and exist over the entire coexistence range as long as the Hopf state is dominant over the Turing ($\gamma_T<\gamma<\gamma_N$), as shown in Figs.~\ref{Fig-multiple} and~\ref{Fig-yTwidth}. 
		
		In the context of 2:1 frequency locking, the comb-like states correspond to spatially localized resonances that further extend the frequency locking regime outside the resonance tongue. Notably, localized comb-like states have been also observed in vibrating granular media and referred to as ``decorated fronts''~\cite{Granular}. However, in these experiments they seem to form near resonant domain patterns and thus, their formation mechanisms may be unrelated to the Hopf-Turing bifurcation.
		
		To this end, using the generic amplitude equation framework, we have presented a selection mechanism that allows us to understand and robustly design spatially localized reaction-diffusion patterns in two dimensional geometries~\cite{vanag2008design,dewel1995pattern,borckmans1995localized}. Specifically, these results indicate the origin of intriguing spiral waves with stationary cores that have been observed in CIMA~\cite{kepper,FlipFlopand2Dspirals} and suggest the formation of localized resonant patterns outside the classical 2:1 frequency locking region, as such in the case of periodically driven Belousov-Zhabotinski chemical reaction~\cite{ArikLabyrinthine}. {A detailed analysis/comparison with reaction-diffusion models for chemical reactions is however, beyond the scope of this work and should be addressed in future studies.}
		
		
		\begin{acknowledgments} 
			We thank Ehud Meron and Francois A. Leyvraz for fruitful discussions, and P.M.C. also acknowledges the use of Miztli supercomputer of UNAM under project number LANCAD-UNAM-DGTIC-016. This work was supported by the Adelis Foundation, CONACyT under project number 219993, UNAM under projects DGAPA-PAPIIT IN100616 and IN103017.  
		\end{acknowledgments}
		

		
		%

\end{document}